\documentclass[numberedappendix]{emulateapj}

\usepackage{epsfig, natbib, graphicx, color}
\tightenlines

\countdef\decade=200
\advance\decade by \year
\countdef\hours=201	
\advance\hours by \time
\divide\hours by 60
\countdef\mins=202
\advance\mins by \hours
\multiply\mins by 60
\multiply\hours by 100
\countdef\miltime=203
\advance\miltime by \hours
\advance\miltime by \time
\advance\miltime by -\mins

\newcommand{\ltsim}{\protect\raisebox{-0.5ex}{$\:\stackrel{\textstyle <}
        {\sim}\:$}}
\newcommand{\gtsim}{\protect\raisebox{-0.5ex}{$\:\stackrel{\textstyle >}
        {\sim}\:$}}

\newcommand{\hii}{H~\textsc{ii}\ }
\newcommand{\krho}{k_{\rho}}
\newcommand{\epsth}{\epsilon_0}
\newcommand{\rii}{r_{\rm II}}
\newcommand{\rhoii}{\rho_{\rm II}}
\newcommand{\cii}{c_{\rm II}}
\newcommand{\tii}{T_{\rm II}}
\newcommand{\uii}{u_{\rm II}}
\newcommand{\muii}{\mu_{\rm II}}
\newcommand{\xii}{x_{\rm II}}

\newcommand{\alphab}{\alpha_{\rm B}}
\newcommand{\riidot}{\dot{r}_{\rm II}}
\newcommand{\msh}{M_{\rm sh}}
\newcommand{\ash}{A_{\rm sh}}
\newcommand{\kb}{k_{\rm B}}
\newcommand{\msun}{M_{\odot}}
\newcommand{\ftrapIR}{f_{\rm trap,IR} }
\newcommand{\Teffsh}{T_{\rm eff,sh}}
\newcommand{\avir}{\alpha_{\rm vir}}

\begin{document} 

\title{The Dynamics of Radiation Pressure-Dominated \hii Regions}

\slugcomment{Accepted for publication in ApJ}

\author{Mark R. Krumholz}
\affil{Department of Astronomy \& Astrophysics, University of California, Santa Cruz, CA 95064}
\email{krumholz@ucolick.org}

\author{Christopher D. Matzner}
\affil{Department of Astronomy, University of Toronto, Toronto, ON, M5S 3H8, Canada}

\begin{abstract}
We evaluate the role of radiation pressure in the dynamics of \hii regions. We first determine under what conditions radiation pressure is significant in comparison to gas pressure and show that, while radiation pressure is generally unimportant for \hii regions driven by a handful of massive stars, it is dominant for the larger \hii regions produced by the massive star clusters found near the Galactic center and in starburst environments. We then provide a solution for the problem of how \hii regions expand when radiation pressure influences their behavior. Finally, we compare radiation-dominated \hii regions to other sources of stellar feedback, and argue that \hii regions are probably the primary mechanism for regulating the formation of massive star clusters.
\end{abstract}


\section{Introduction}
\label{intro}

The dynamical expansion of a bubble of warm, ionized gas produced by a young star or star cluster is one of the classical problems of the interstellar medium. \hii region dynamics are important in part because they play a dominant role in regulating the formation of star clusters. \citet{williams97} and \citet{carpenter00a} argue based on observations that no more than $\sim 10\%$ of the mass in a giant molecular cloud can ever be incorporated into stars, and \citet{matzner02} and \citet{krumholz06d} present GMC models that explain this inefficiency quantitatively in terms of the evaporation of GMC mass by \hii regions. Nearby star clusters provide direct evidence for this phenomenon. For example in the Orion Nebula Cluster, the radiation of $\theta^1$ Ori C launches an ionized wind from the molecular cloud surface that carries a mass flux of $\sim 10^{-2}$ $\msun$ yr$^{-1}$ \citep{Odell01a}, sufficient to ablate a mass comparable to the ONC's stellar mass in $\sim 10^5$ yr. Any understanding of star formation rates and efficiencies, and of molecular cloud lifetimes, must therefore be based on an understanding of \hii regions.

Most work on this problem to date has focused on \hii regions where conditions are similar to those found in the Galaxy within several kpc of the Sun. The star clusters in this region are typically below $10^4$ $\msun$ in mass \citep[e.g.][]{williams97, lada03}, and since a fully-sampled IMF at zero age produces $S \sim 10^{46.5}$ ionizing photons s$^{-1}$ per $\msun$ of stars \citep{krumholz06d}, they have luminosities of at most $S\sim 10^{50}$ ionizing photons s$^{-1}$. For ionizing luminosities in this range, both numerical treatments and analytic estimates show that radiation pressure is generally small compared to gas pressure inside \hii regions \citep[e.g.][]{mathews69a, gail79a, arthur04a, henney07a}. Thus while radiation on dust grains may play an important role producing small holes in \hii region centers \citep[e.g.][]{mathews67a, inoue02a}, the standard assumption has been that radiation pressure is not significant for determining the dynamics the \hii region as a whole. Treatments of \hii regions under circumstances where radiation pressure is significant have generally been limited to hydrostatic models that do not include any dynamics \citep[e.g.][]{dopita02a, dopita03a, dopita06a} or numerical models of particular regions \citep[e.g.][]{pellegrini07a, pellegrini09a, harper-clark09a}. The former provide little information on how radiation pressure affects gas motion, while the latter do not easily lead to general conclusions about when radiation pressure is important and how \hii region dynamics are altered when it is.

A re-assessment of the role of radiation pressure in \hii region dynamics is timely because observations of star cluster formation are beginning to probe new regimes where earlier arguments minimizing the role of radiation pressure no longer apply. The importance of radiation pressure rises as the ionizing luminosity does, and in contrast to the values of at most $S\sim 10^{50}$ s$^{-1}$ found in local \hii regions, the Arches cluster near the Galactic center has an ionizing luminosity of $S\approx 4\times 10^{52}$ s$^{-1}$ \citep{figer02a}, and some extragalactic clusters have even larger values.

Moreover, regions with large ionizing luminosities also tend to have large virial and escape velocities, and in this case the conventional description of \hii region expansion driven by ionized gas pressure also breaks down.
\citet{mccrady07a} measure velocity dispersions of $10-30$ km s$^{-1}$ for the super star clusters in M82, and this probably represents a lower limit on the velocity dispersions and escape velocities in the clusters' parent molecular clouds. Since the sound speed in ionized gas is only $10$ km s$^{-1}$, an \hii region driven by gas pressure could not have expanded within or driven mass out of the the clouds from which these star clusters formed. We are forced to conclude that either these clusters must have formed with an efficiency of nearly 100\%, or that some mechanism other than ionized gas pressure is responsible for removing mass and limiting the efficiency.

While we cannot directly rule out the hypothesis of 100\% star formation efficiency for the M82 clusters, we can rule it out for similar large clusters in other galaxies. For example, the young star clusters observed in the  Antennae galaxies have comparable escape velocities \citep{whitmore99a}, but \citet{fall05a} and \citet{whitmore07a} find that the star clusters in the Antennae show the same ``infant mortality" phenomenon as those in the Milky Way: roughly 90\% of all star clusters dissolve within $\sim 10$ Myr of formation, almost certainly because the formation process operates with a low efficiency and removal of a majority of gas leaves the remaining stars unbound. Clearly some mechanism must remove gas from these clusters as they form, and we argue below that radiation-driven \hii regions are the most natural explanation.

Our approach to the problem is as follows. In \S~\ref{derivation} we derive a condition for when radiation pressure is important, and we then give a solution to the idealized problem of an \hii region expanding into an ambient medium including radiation pressure effects. In \S~\ref{trapping} we discuss the contribution of trapped radiation to \hii region dynamics. In \S~\ref{supernovae} we discuss the relative importance of radiation-driven \hii regions and supernovae. Finally, we summarize in \S~\ref{summary}.

\section{The Effects of Radiation Pressure}
\label{derivation}

\subsection{When is Radiation Pressure Significant?}

\begin{figure}
\plotone{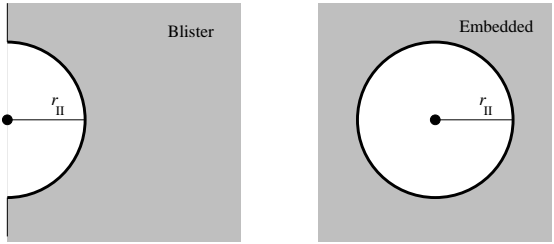}
\caption{
\label{diagram_hiireg}
Schematic representation of the blister ({\it left}) and embedded ({\it right}) \hii region cases. The gray represents in the ambient cloud, and the black dot is the driving source. In the blister case the \hii region is hemispherical, while in the embedded case it is spherical.
}
\end{figure}

Consider a source of bolometric luminosity $L$ that produces ionizing photons at a rate $S$, located at $r=0$. Following \citet{matzner02}, we investigate two cases, which can be treated in parallel. The first is a region of neutral gas of density $\rho=\rho_0 (r/r_0)^{-\krho}$. The second is a region in which the density is $\rho=\rho_0 (r/r_0)^{-\krho}$ for $x>0$ and 0 for $x<0$. The former corresponds to the case of an ``embedded" \hii region that is completely surrounded by dense gas, and the latter to a ``blister" \hii region in which the driving source is at the edge of a dense cloud, and the ionized gas can escape freely. We illustrate these two possible configurations in Figure \ref{diagram_hiireg}. We define $\epsth = 13.6$ eV as the threshold photon energy required to ionize a neutral hydrogen atom, and for convenience we define $\psi=L/(S\epsth)$ to be the ratio of the star's bolometric power to its ionizing power, counting only an energy $\epsth$ per ionizing photon. For massive stars and clusters whose luminosity comes mostly from massive stars, $\psi\sim 1$.

We wish to determine how the gas moves in response to the radiation flux, and to understand under what circumstances radiation plays an important role in determining gas motions. Following the usual procedure, we approximate that the ionized gas is isothermal at temperature $\tii$, and has a sound speed $\cii$ which is much larger than the sound speed in the neutral gas. If radiation pressure is negligible, we have the usual \citet{spitzer78} solution: the ionized material expands due to its thermal pressure. This expansion sweeps the neutral gas into a thin shell, which contains most of the mass that was originally inside the radius $\rii$ of the \hii region. If radiation pressure is the dominant force acting on the gas then a fluid element at a distance $r$ from the source undergoes a radiative acceleration $a_{\rm rad} = \int\, \kappa_\nu(r) L e^{-\tau_\nu(r)}/(4 \pi r^2 c)\, d\nu$, where $\kappa_\nu(r)$ is the opacity of the fluid at radius $r$ to photons of frequency $\nu$, and $\tau_\nu(r) = \int_{0}^r \kappa_\nu(r') \rho(r') \, dr'$ is the optical depth from the source to that point. Photons below the Lyman limit carry roughly half the radiative momentum, and since these are absorbed primarily by dust grains, $\kappa_\nu(r) e^{-\tau_\nu(r)}/r^2$ and thus the radiative acceleration is a decreasing function of $r$. The other half of the momentum is carried by photons above the Lyman limit, which can be absorbed by either H~\textsc{i} or dust grains -- see Appendix \ref{dustabsorption}. If dust absorption dominates, then radiative acceleration falls with radius as for lower-energy photons. If H~\textsc{i} absorption dominates, the acceleration is proportional to the recombination rate, which is either flat (if the gas density is uniform) or again declines with radius (if radiative acceleration causes gas to pile up near the shell edge). Thus, the total radiative acceleration is always largest closest to the source, and again material will be swept into a thin shell of radius $\rii$. The interior of this shell will be optically thin.

Thus, after a rapid initial expansion phase, the dynamics of the gas reduce to the problem of computing the dynamics of the thin shell that bounds the \hii region. Following \citet{matzner02}, we can solve this problem by writing down the momentum equation for the shell:
\begin{equation}
\label{momeqn}
\frac{d}{dt} (\msh \riidot) = \ash \left\{\rhoii [\cii^2 + \uii (\uii - \riidot)] + \frac{f_{\rm trap} L}{4\pi \rii^2 c}\right\},
\end{equation}
where $\msh$ and $\ash$ are the shell's mass and area, and $\rhoii$ and $\uii$ are the density and velocity of the gas immediately interior to it. The shell area and mass are $\ash = (4,2)\pi \rii^2$ and $\msh = (4,2)\pi \rii^3 \overline{\rho}/3$, where $\overline{\rho}(r)=[3/(3-\krho)]\rho_0 (r/r_0)^{-\krho}$ is the mean density inside the spherical or hemispherical region of radius $r$ in the initial cloud, and the values in parentheses refer to the cases for a (spherical, hemispherical) \hii region. The first term on the right-hand side represents gas pressure, while the second represents radiation pressure. Note that in writing this equation we have implicitly assumed that all the radiation force is applied at the thin shell, rather than in the \hii region interior. This is certainly a good approximation when radiation pressure is dominant, since, as discussed above, the interior of the shell will be cleared by radiation pressure, and thus all photons will be absorbed in or near the shell. When gas pressure dominates and the shell interior is of uniform density, the rate of momentum deposition by ionizing photons matches the recombination rate. Since this is uniform, so the mean radius at which momentum is deposited is $3/4$ of the shell radius. Moreover, non-ionizing photons, which carry half the total momentum, still deposit all their momentum in the shell. Thus our approximation that all the momentum is deposited in the shell is a good one.

The quantity $f_{\rm trap}$ in equation (\ref{momeqn}) represents the factor by which the radiation pressure force is enhanced by trapping of energy within the expanding shell. If $f_{\rm trap} = 0$, then the shell is optically thin and all stellar photons escape without depositing any momentum. Given the high color temperature of the emitting stars and the large opacity of the gas clouds where massive clusters form, this is not realistic. A value $f_{\rm trap} = 1$ corresponds to every photon emitted by the stars being absorbed once in the shell and depositing its momentum there, then promptly escaping. If there is more than one interaction per photon then $f_{\rm trap}$ could potentially be much larger than unity. In this case the velocity to which the shell accelerates will be limited by the rate at which the stars supply energy rather than momentum.

Trapping can happen in three ways. First, some fraction of the stellar radiation will go into accelerating line driven winds off stellar surfaces \citep{castor75}, and the expanding wind will collide with the slower-moving shell. This will produce some transfer of momentum, which could be large if the shocked gas becomes trapped inside the shell \citep{castor75a, weaver77a}. Second, if the shell is sufficiently optically thick to long-wavelength radiation, then ultraviolet and visible photons that are absorbed by dust grains in the shell and re-radiated at infrared wavelengths may remain trapped in the shell and interact more times before finally escaping. Third, Lyman $\alpha$ photons that are produced by recombinations in the shell or in the \hii region interior may undergo many resonant interactions before escaping. We defer a discussion of these trapping mechanisms until \S~\ref{trapping}, and for now we simply assert the result from that section: $f_{\rm trap}$ is always likely to be of order a few. For this reason, we choose to leave it as a free parameter of constant value, for which we adopt a fiducial value $f_{\rm trap}=2$ when we wish to perform numerical evaluations.

We can characterize when radiation pressure is significant by examining the limiting cases of gas- and radiation-pressure dominated flows. In the gas-pressure dominated case, we have the usual \citet{spitzer78} \hii region solution. Once expansion of the \hii region becomes subsonic with respect to the ionized gas, the \hii region interior approaches a uniform density, and ionization balance requires that
\begin{equation}
\label{rhoiieqn}
\frac{4}{3}\pi \rii^3 \alphab \left(\frac{\rhoii \cii^2}{\kb \tii}\right)^2 = \phi S
\end{equation}
where $\rii$, $\tii$, and $\cii$ are the radius, temperature, and sound speed of the ionized region, $\alphab$ is the case-B recombination coefficient, $\phi$ is a dimensionless number that accounts for absorption of ionizing photons by dust grains and for free electrons provided by elements other than hydrogen, and we have adopted the usual on-the-spot approximation. If He is singly-ionized and 27\% of photons are absorbed by dust rather than gas, as expected for gas pressure-dominated \hii regions with Milky Way dust-to-gas ratios \citep{mckee97}, then $\phi=0.73$. We discuss the value of $\phi$, and of dust absorption generally, in more detail in Appendix \ref{dustabsorption}. Note that equation (\ref{rhoiieqn}) holds approximately even in the case of a blister-type hemispherical \hii region. Following \citet{matzner02}, we consider in that in the case of an embedded, spherical \hii region that $\uii \ll \cii$, while for a blister-type one $\uii\approx 2\cii$. Thus the gas pressure term on the right-hand side of equation (\ref{momeqn}) becomes
\begin{equation}
\rhoii [\cii^2 + \uii (\uii - \riidot)] \approx (1,2) \sqrt{\frac{3 S \phi}{4\pi \alphab}} \frac{\kb\tii}{\rii^{3/2}}.
\end{equation}

In the limiting case of a radiation-pressure dominated flow, the radiation pressure term is simply $L/(4\pi \rii^2 c)$. Since the radiation- and gas-pressure terms have different radial dependences, we can calculate a characteristic radius for which they are equal:
\begin{eqnarray}
r_{\rm ch} & = & \frac{\alphab}{12 (1,4) \pi \phi} \left(\frac{\epsth}{\kb\tii}\right)^2 f_{\rm trap}^2 \frac{\psi^2 S}{c^2} \\
& \rightarrow & (9.2, 2.3)\times 10^{-2} \,  S_{49}\mbox{ pc},
\label{rchnum}
\end{eqnarray}
where the numerical evaluation is for our fiducial parameters $\tii=7000$ K, $\phi=0.73$, $f_{\rm trap}=2$, $\psi=1$, and $\alphab=3.46\times 10^{-13}$ cm$^3$ s$^{-1}$, and $S_{49}=S/10^{49}$ s$^{-1}$. Since radiation forces vary with radius as $\rii^{-2}$, radiation dominates at smaller radii and gas pressure at larger radii.

It is useful to compare this to the Str\"{o}mgren radius at which gas pressure-driven expansion begins in the case of negligible radiation pressure force. Setting $\rhoii$ in equation (\ref{rhoiieqn}) equal to the mean density $\overline{\rho}(r_{\rm St,0})$ inside $r_{\rm St,0}$, we find
\begin{equation}
r_{\rm St,0}  = \left(\frac{3\phi S}{4\pi \alphab}\right)^{1/3} \left(\frac{\kb \tii}{\overline{\rho}(r_{\rm St,0})\cii^2}\right)^{2/3},
\end{equation}
and computing the ratio of $r_{\rm ch}$ to this we find
\begin{eqnarray}
\label{zeta}
\zeta & \equiv & \frac{r_{\rm ch}}{r_{\rm St,0}} =
f_{\rm trap}^2 \left(\frac{\epsth \psi}{(1,2)\kb\tii c}\right)^2 
\left(\frac{\alphab^2 \overline{\rho}(r_{\rm St,0}) S}{36\pi \phi^2 \muii m_{\rm H}}\right)^{2/3}
\\
& \rightarrow & (6.2, 1.6)\times 10^{-2} \, (\overline{n}_{\rm H,2} S_{49})^{2/3},
\end{eqnarray}
where $\muii=0.61$ is the mean molecular weight in the fully ionized gas (so $\cii=9.74$ km s$^{-1}$) and $\overline{n}_{\rm H,2}=\overline{\rho}(r_{\rm St,0})/(100 \mu m_{\rm H})$ is the mean density of H nuclei in units of $10^2$ cm$^{-3}$, and $\mu=1.4$ is the atomic mass per H nucleus for gas of standard cosmic composition.

\begin{deluxetable*}{lcccccccc}
\tablecaption{Sample of star clusters\label{clusterlist}}
\tablehead{
\colhead{Name} &
\colhead{$M$ ($10^5$ $\msun$)} &
\colhead{$R$ (pc)} &
\colhead{$\log \overline{n}_{\rm H,2}$} &
\colhead{$S_{49}$} &
\colhead{$r_{\rm ch}$ (pc)} &
\colhead{$\zeta$} &
\colhead{$r_{\rm stall}$ (pc)} &
\colhead{Ref.}
}
\startdata
ONC & 0.046 & 0.8 & 2.8 & 2.7 & 0.062 & 2.2 & 0.94 & 1, 2 \\
Arches & 0.2 & 0.4 & 4.3 & 400 & 9.2 & 650 & 0.40 & 3, 4 \\ 
NGC 5253 & 5 & 1.0 & 4.5 & 4000 & 92 & $30000$ & 0.55 & 5 \\
M82 L\tablenotemark{a} & 40 & 1.4 & 5.0 & 13000\tablenotemark{b} & 290 & $17000$ & 0.44 & 6 \\
M82 F & $5.5$ & 1.5 & 4.0 & 1700\tablenotemark{b}  & 40 & 1100 & 0.82 & 6 \\
M82 11 & $3.9$ & 1.1 & 4.3 & 1200\tablenotemark{b} & 28 & 1200 & 0.57& 6 \\
M82 9 & $23$ & 2.5 & 4.0 & 7300\tablenotemark{b} & 170 & 2700 & 1.2 & 6 \\
M82 8 & $4.0$ & 1.6 & 3.9 & 1300\tablenotemark{b}  & 29 & 680& 0.91& 6 \\
M82 7 & $22$ & 2.7 & 3.9 & 7000\tablenotemark{b}  & 160 & 2200& 1.4 & 6 \\
M82 6 & $2.7$ & 1.4  & 3.9 & 850\tablenotemark{b}  & 20 & 520& 0.83 & 6 \\
M82 k & $5.7$  & 3.0 & 3.2 & 1800\tablenotemark{b} & 41 & 290 & 2.3 & 6 \\
M82 m & $7.3$ & 1.4 & 4.3 & 2300\tablenotemark{b} & 53 & 2000 & 0.62 & 6 \\
M82 q & $2.8$ & 1.9 & 3.4 & 890\tablenotemark{b}  & 20 & 280& 1.4& 6 \\
M82 3 & $2.7$ & 1.5 & 3.7 & 850\tablenotemark{b}  & 20 & 420& 1.0 & 6 \\
M82 1a & $8.6 $ & 2.1 & 3.8 & 2700\tablenotemark{b}  & 62 & 1100& 1.1 & 6 \\
M82 1c & $5.2$ & 1.5 & 4.0 & 1600\tablenotemark{b} & 38 & 1000 & 0.80 & 6 \\
M82 r & $3.0$ & 1.7  & 3.6 & 950\tablenotemark{b}  & 22 & 370& 1.2 & 6 \\
M82 t & $2.5$ & 1.7 & 3.5 & 790\tablenotemark{b}  & 18 & 290& 1.2 & 6 \\
Ant B1\tablenotemark{c} & 42 & 29\tablenotemark{d}  & 1.0 & 5100 & 120 & 24 & 34 & 7 \\
Ant B & 50 & 49\tablenotemark{d}  & 0.47 & 5500 & 130 & 9.9 & 71 & 7 \\
Ant D & 19 & 40\tablenotemark{d}  & 0.30 & 1500 & 36 & 3.3 & 68 & 7 \\
Ant C & 41 & 21\tablenotemark{d}  & 1.5 & 1300 & 30 & 18 & 16 & 7 \\
Ant D1 & 16 & 11\tablenotemark{d}  & 2.0 & 100 & 2.3 & 6.9 & 5.2 & 7 \\
Ant D2 & 8.0 & 35\tablenotemark{d}  & 0.11 & 1000 & 23 & 1.8 & 79 & 7 \\
Ant E1 & 2.6 & 13\tablenotemark{d}  & 0.91 & 170 & 3.8 & 1.9 & 21 & 7 \\
Ant E2 & 4.1 & 25\tablenotemark{d}  & 0.25 & 170 & 3.9 & 0.7 & 48 & 7 \\
Ant E3 & 0.7 & 11\tablenotemark{d}  & 0.60 & 170 & 3.9 & 1.2 & 31 & 7 \\
Ant F & 7.4 & 9.3\tablenotemark{d}  & 1.8 & 220 & 5.0 & 8.9 & 7.3 & 7 \\
Ant E5 & 26 & 23\tablenotemark{d}  & 1.2 & 200 & 4.7 & 3.2 & 16 & 7 \\
Ant F2 & 3.5 & 17\tablenotemark{d}  & 0.71 & 440 & 10 & 2.7 & 32 & 7 \\
Ant F1 & 15 & 15\tablenotemark{d}  & 1.4 & 74 & 1.7 & 2.5 & 9.2 & 7 \\
Ant E4 & 6.5 & 36\tablenotemark{d}  & -0.01 & 1100 & 26 & 1.6 & 94 & 7 \\
Ant A1 & 5.0 & 15\tablenotemark{d}  & 1.0 & 2100 & 48 & 12 & 30 & 7 \\
Ant S & 32 & 14\tablenotemark{d}  & 1.9 & 210 & 3.9 & 10 & 6.3  & 7 \\
\enddata
\tablenotetext{a}{M82 objects are super star clusters, whose identifiers correspond to those given in \citet{mccrady07a}}
\tablenotetext{b}{An ionizing luminosity for this object has not been reported in the literature, so we estimate it by taking  $S_{49} = 10^{-2.5} (M/\msun)$, the value for a fully-sampled zero-age IMF \citep{krumholz06d}.}
\tablenotetext{c}{Antennae objects are emission line clusters, whose identifiers correspond to those given in \citet{gilbert07a}}
\tablenotetext{d}{This radius is almost certainly an overestimate of the radius at formation, since the Antennae emission line clusters have undergone significant expansion \citep{gilbert07a}. Correcting for this effect would raise $\overline{n}_{\rm H,2}$, which in turn would increase $\zeta$ and decrease $r_{\rm stall}$.}
\tablecomments{Col.\ (2): Stellar mass. Col.\ (3): Half-light radius, except for Antennae objects, where we use $1/2$ of FWHM. Col.\ (4): Density, computed as $\overline{n}_{\rm H,2} =[3 M/(4\pi R^3)]/(100 \mu m_H)$. Col.\ (5): Ionizing luminosity. Col.\ (6-8): $r_{\rm ch}$, $\zeta$, and $r_{\rm stall}$, computed with fiducial parameters in the blister case with $\krho=0$ (equations \ref{rchnum}, \ref{zeta}, and \ref{rstall}).}
\tablerefs{1 = \citet{hillenbrand98}, 2 = \citet{williams97}, 3 = \citet{figer99a}, 4 = \citet{kim00a}, 5 = \citet{turner00a}, 6 = \citet{mccrady07a}, 7 = \citet{gilbert07a}}
\end{deluxetable*}

\begin{figure}
\plotone{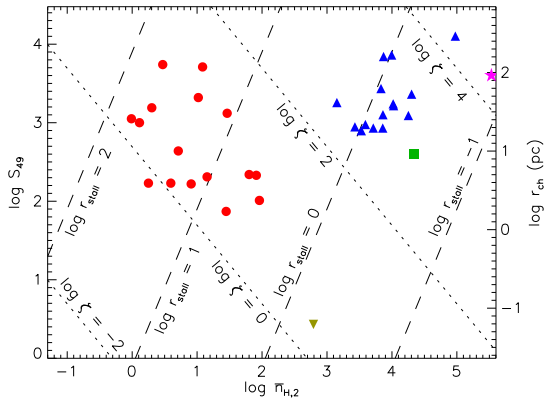}
\caption{
\label{clustersample}
$S_{49}$ versus $\overline{n}_{\rm H,2}$ for a sample of massive star clusters. The value of $r_{\rm ch}$ shown on the right axis is derived from equation (\ref{rchnum}), and the lines of constant $\zeta$ ({\it dotted}) and $r_{\rm stall}$ ({\it dashed}) are derived using equations (\ref{zeta}) and (\ref{rstall}). All are for the blister case. Values of $r_{\rm stall}$ indicated in the figure are in pc. The data shown are described in Table \ref{clusterlist}: emission line clusters in the Antennae Galaxies ({\it red circles}), super star clusters (SSCs) in M82 ({\it blue triangles}), the SSC in NGC 5253 ({\it purple star}), and the Arches ({\it green square}) and Orion Nebula ({\it brown inverted triangle}) Clusters in the Milky Way. 
}
\end{figure}

Thus we see that for single OB stars, $S_{49}\sim 1$, expanding into Galactic molecular clouds, $\overline{n}_{\rm H,2}\sim 1$, radiation force is negligible once the \hii region reaches a tenth of a pc in size ($r_{\rm ch}\la 0.1$ pc), and is often negligible as soon as the ionized region has finished its initial rapid expansion to the Str\"{o}mgren radius ($\zeta \la 1$). Thereafter the usual gas pressure-driven expansion solution applies. However, we reach a very different conclusion if we consider the formation of very massive clusters in dense environments. In Table \ref{clusterlist} we list properties for a sample of massive star clusters in the Milky Way, M82, the Antennae, and NGC 5253. We plot  $S_{49}$ versus $\overline{n}_{\rm H,2}$ for these objects in Figure \ref{clustersample}. As the plot shows, these clusters have $\zeta$ in the range $\sim 1 - 10^4$, indicating that they go from the border between radiation- and gas-dominated to completely radiation-dominated. The characteristic radii where gas pressure becomes comparable to radiation pressure run from $\sim 1-100$ pc, which is generally larger than the physical size of the cluster in question. The exception is the Orion Nebular Cluster (the inverted triangle), by far the smallest cluster shown in Figure \ref{clustersample}. This has $r_{\rm ch}=0.06$ pc (assuming a blister-type \hii region, which is observed), considerably smaller than the $0.8$ pc radius of the cluster.

We should add a caution here, which is that for the clusters described in Table \ref{clusterlist} and plotted in Figure \ref{clustersample}, we have computed $r_{\rm ch}$, $\zeta$, and $r_{\rm stall}$ using the full ionizing luminosity of the cluster. This is not correct very early in the expansion process, since the initial Str\"{o}mgren radius around each massive star is so small that it may enclose at most a few of its neighbors. Since this will lower the ionizing luminosity compared to our value, the very early expansion could be gas-driven. However, as the expanding shells overlap, the incorporation of more stars into their interiors will rapidly convert the expansion to a radiatively-driven one; since $\zeta \propto S_{49}^{2/3}$, the expanding shell in a cluster for which we have computed $\zeta = 100$ reaches $\zeta > 1$ and becomes radiation-dominated when the shell includes only $4.6\%$ of the cluster luminosity. Since luminosity tends to be strongly centrally concentrated, this will occur well before the shell includes this fraction of the cloud volume. Thus radiation takes over very early in the expansion process for those clusters for which we have computed values of $\zeta \gg 1$.

\subsection{The Dynamics of Expansion with Radiation Pressure}

To study the dynamics of an expanding \hii region with significant radiation pressure, it is convenient to non-dimensionalize the equation of motion (\ref{momeqn}). We let $x=r/r_{\rm ch}$ and $\tau=t/t_{\rm ch}$, where
\begin{equation}
t_{\rm ch} = \sqrt{\frac{4\pi}{3-\krho} \frac{\rho_0 r_0^{\krho} c}{f_{\rm trap} L} r_{\rm ch}^{4-\krho}}.
\end{equation}
With this change of variables equation (\ref{momeqn}) becomes
\begin{equation}
\label{momeqnnondim}
\frac{d}{d\tau}\left(\xii^{3-\krho} \frac{d}{d\tau}\xii\right) = 1 + \xii^{1/2}.
\end{equation}
The first term on the right hand side represents radiation pressure, and the second represents gas pressure. This equation is not exactly correct near $\xii=1$, because in writing the gas pressure term we have implicitly assumed that the density inside the \hii region is uniform and has the value given by ionization balance, equation (\ref{rhoiieqn}). This assumption cannot be precisely true when $\xii\ltsim 1$, because while radiation pressure is significant it will exert a force on material in the \hii region interior that will make the density higher toward the shell wall than near the \hii region center, while equation (\ref{rhoiieqn}) assumes uniform density. Nonetheless, equation (\ref{momeqnnondim}) represents a reasonable approximation that becomes exact everywhere except near $\xii=1$.

\begin{figure*}
\plotone{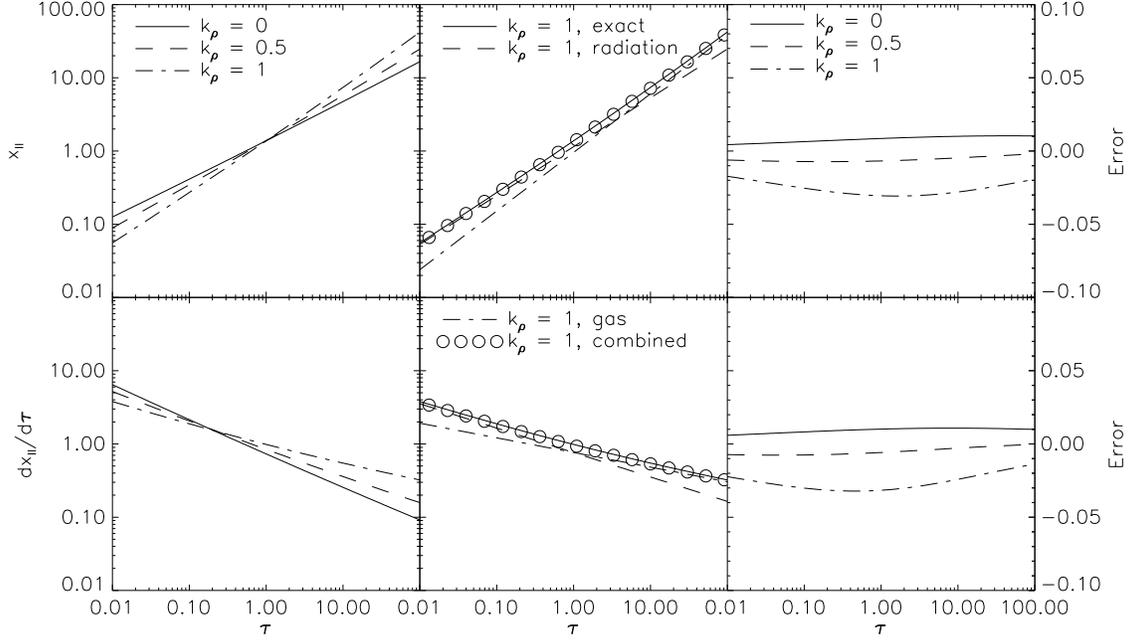}
\caption{
\label{xiisol}
Dimensionless radius $\xii$ ({\it top row}) and velocity $d\xii/d\tau$ ({\it bottom row}) versus time $\tau$. In the left column we show the numerical solutions for $\krho=0$, $0.5$, and $1$. In the middle we show a comparison of the numerical solution and with the similarity solutions for pure radiation ({\it dashed line}, equation \ref{radsolution}), pure gas pressure ({\it dot-dashed line}, equation \ref{gassolution}), and an approximation that combines them ({\it circles}, equation \ref{approxsolution}). In the right column we show the error in our approximate solution (equation \ref{approxsolution}), defined as the difference between the approximate and numerical solutions, normalized to the numerical solution.
}
\end{figure*}

At early times, when $\xii \ll 1$, we can drop the gas pressure term $\xii^{1/2}$ on the right hand side, and the resulting equation admits the similarity solution 
\begin{equation}
\label{radsolution}
x_{\rm II,rad} = \left(\frac{4-\krho}{2} \tau^2\right)^{1/(4-\krho)}.
\end{equation}
The dynamics of this solution can be understood by noting that since the momentum of the shell equals the radiant momentum modified by the trapping factor ($\msh \riidot = f_{\rm trap} Lt/c$), the shell's kinetic energy $\msh \riidot^2/2$ is a very small fraction, $f_{\rm trap}^2 \riidot/(2c)$, of the total radiated energy $L t$.  In fact, if $f_{\rm trap}\sim 1$, then the energy of the shell's motion approximately matches the energy of the photons currently crossing its interior: $\msh\riidot^2/2 \approx (\eta/2) L \rii/c$, where $\eta=\riidot t/\rii = 2/(4-\krho)$.    The inefficiency of direct radiation driving is related to the low value of $f_{\rm trap}$, an issue we return to in \S~\ref{trapping}.

If we were to drop the radiation pressure term, we would have the usual similarity solution for gas pressure expansion
\begin{equation}
\label{gassolution}
x_{\rm II,gas} = \left[\frac{(7-2\krho)^2}{4(9-2\krho)} \tau^2\right]^{2/(7-2\krho)}.
\end{equation}
Although the exact solution will approach this value when $\xii\gg 1$, the existence of a characteristic scale $r_{\rm ch}$ implies that there is no true similarity solution that includes both the radiation- and gas-driven phases. When radiation pressure is significant it provides an extra boost of momentum in the phase before gas pressure takes over, accelerating the expansion relative to the standard similarity solution. This breaks the self-similarity of the gas-driven expansion phase, although the expansion approaches self-similarity as $\tau\rightarrow \infty$ and the extra momentum provided by the radiation becomes small compared to that input by the gas. In the absence of a similarity solution, however, it is trivial to integrate equation (\ref{momeqnnondim}) numerically. We do so subject to the boundary condition that $\xii$ and $d\xii/d\tau$ approach the values that correspond to the similarity solution (\ref{radsolution}) as $\tau\rightarrow 0$, and plot the result in Figure \ref{xiisol} for some sample values of $\krho$. For comparison, we also show the pure radiation and pure gas similarity solutions. The true solution may be reasonably approximated by an appropriately weighted sum between the two. The expression
\begin{equation}
\label{approxsolution}
x_{\rm II,approx} = 
\left(x_{\rm II,rad}^{(7-\krho)/2} + x_{\rm II,gas}^{(7-\krho)/2}\right)^{2/(7-\krho)}
\end{equation}
is accurate to better than 5\% for $\krho=0-1$.

\subsection{\hii Region Stalling}

As discussed in \S~\ref{intro}, one of the reasons that radiation-driven expansion is of interest is that gas-driven expansion will fail in regions where the ambient velocity dispersion and escape velocity exceed the ionized gas sound speed \citep[e.g.][]{matzner02, krumholz06d}. The expansion velocity is $\riidot=v_{\rm ch} (d\xii/d\tau)$, where
\begin{equation}
v_{\rm ch} =\frac{r_{\rm ch}}{t_{\rm ch}} = \sqrt{(3,8)}\zeta^{(2\krho-3)/4} \cii.
\end{equation}
The expansion rate therefore drops to the ionized gas sound speed when $d\xii/d\tau = \zeta^{(3-2\krho)/4}/\sqrt{(3,8)}$. Since we have solved for $d\xii/d\tau$ numerically, it is trivial to numerically invert this equation for a given $\zeta$ to obtain the radius and time at which the expansion becomes subsonic. However, it is more illuminating to consider the radiation-dominated case $\zeta\gg 1$, because in that case the similarity solution (\ref{radsolution}) applies, and the resulting analytic form for $d\xii/d\tau$ allows us to solve for the radius at which the subsonic transition occurs in closed form. In this limit, expansion becomes subsonic at
\begin{equation}
x_{\rm sub} = 
\left[\frac{4-\krho}{2(3,8)}\zeta^{(3-2\krho)/2}\right]^{1/(\krho-2)}.
\end{equation}
To give some idea of this in physical units, for $\krho=0$ the radius at which the expansion becomes subsonic is
\begin{equation}
\label{rsub}
r_{\rm sub} = (0.90, 0.90) (S_{49}/\overline{n}_{\rm H,2})^{1/2}\mbox{ pc}.
\end{equation}
Clearly the transition to subsonic expansion will not happen until an \hii region has swept up a significant fraction of the gas in the protocluster, and it is therefore likely that radiation-driven \hii regions can expel gas from clouds where gas-driven ones could not.

Alternately, we can ask when the \hii region expansion velocity becomes comparable to the velocity dispersion within the parent cloud. This should be a reasonable estimate of when the expansion will stall. To the extent that the parent cloud is in virial balance, this will also a good estimate of at what radius the expansion velocity will be reduced to the point where gravity can prevent expansion.\footnote{One might expect that comparing radiation forces to gravitational ones would not produce a characteristic radius, since both vary as $1/r^2$. However, this argument is only valid if the gravitational force is dominated by a central point mass. Forming star clusters, however, are dominated by gas mass, and for a gas mass distribution $\rho\propto r^{-\krho}$, the gravitational force varies as $r^{1-\krho}$. Thus it is possible to obtain a characteristic radius at which gravity balances radiation.} We write the velocity dispersion $\sigma$ of material within a radius $r$ of the cloud center as
\begin{eqnarray}
\sigma & = & \left[\frac{\avir G M(<r)}{5 r}\right]^{1/2} \\
& = & \left[\frac{(4,2)\pi}{15} \avir G \overline{\rho}(r_{\rm St,0}) r_{\rm ch}^2 \zeta^{-\krho} x^{2-\krho}\right]^{1/2},
\end{eqnarray}
where $M(<r)$ is the mass within a radius $r$ of the origin, $x=r/r_{\rm ch}$, and $\avir\equiv 5\sigma^2 r/[G M(<r)]$ is the virial ratio of the material inside radius $r$ \citep{bertoldi92}. The factor $(4,2)$ appears because the gas is spherical in the embedded case and hemi-spherical in the blister case. The velocity dispersion can be either larger or smaller than $\cii$ depending on the density and radius: for $\krho=0$, we have $\sigma/\cii = 0.011 (\avir \overline{n}_{\rm H,2})^{1/2} (r/\mbox{pc})$. The expansion velocity of the shell $v_{\rm ch}(d\xii/d\tau)$ is equal to this velocity at the dimensionless radius $\xii$ for which
\begin{equation}
\label{stalleqn}
\frac{d\xii}{d\tau} = \left[\frac{(4,2)\pi \avir G\overline{\rho}(r_{\rm St,0}) r_{\rm ch}^2}{(45,90)\cii^2}\right]^{1/2} \zeta^{(3-4\krho)/4} \xii^{(2-\krho)/2}.
\end{equation}
As with the problem of determining when the expansion becomes subsonic, it is simple to find a numerical solution to this equation for a given choice of $\overline{n}_{\rm H,2}$ and $S_{49}$, but it is more illuminating to consider the radiation-dominated case, for which we can find a solution analytically. If we set $\xii$ equal to the value given by equation (\ref{radsolution}) for the radiation-dominated phase, then the dimensionless radius $x_{\rm stall}$ that satisfies equation (\ref{stalleqn}) is
\begin{equation}
x_{\rm stall} = \left[\frac{(45,180) \cii^2}{2 \pi (4-\krho) \avir G \overline{\rho}(r_{\rm St,0}) r_{\rm ch}^2} \zeta^{(4\krho-3)/2}\right]^{1/(4-2\krho)}.
\end{equation}
Putting this in dimensional terms, for $\krho=0$ and $\avir=1$ the stalling radius is
\begin{equation}
\label{rstall}
r_{\rm stall} = (8.9, 10.6) \overline{n}_{\rm H,2}^{-1/2} S_{49}^{1/4}\mbox{ pc}.
\end{equation}
Numerical evaluations of $r_{\rm stall}$ for our sample clusters are given in Table \ref{clusterlist}. For most of the clusters shown, the stall radius is comparable to or larger than the total cluster radius. This demonstrates that, unlike gas pressure-driven \hii regions, radiation pressure-driven ones can potentially expel gas from the large clouds that form massive clusters. We show lines of constant $r_{\rm stall}$ (computed for the radiation-dominated case), with observed clusters over-plotted, in Figure \ref{clustersample}.

\section{Radiative Trapping}
\label{trapping}

Our derivation of the shell motion in \S~\ref{derivation} assumes that during the radiation-dominated phase shells are momentum driven, in the sense that the force exerted on the shell is the radiation momentum flux multiplied by at most a modest enhancement factor $f_{\rm trap}$. However, if stellar radiation energy becomes trapped in the bubble interior, then the full amount of bolometric energy emitted by the stars can be used to drive expansion, and $f_{\rm trap}\gg 1$. Here we estimate $f_{\rm trap}$ considering three possible sources of trapping: 
\begin{equation}
f_{\rm trap} = 1+f_{\rm trap,w} + f_{\rm trap,IR} + f_{\rm trap,Ly\alpha},
\end{equation}
where the 1 represents absorption of the direct radiation and the remaining three factors on the right are the contributions to trapping due to stellar winds, infrared photons, and Ly$\alpha$ photons.

\subsection{Stellar Winds}
\label{winds}

Massive stars emit strong winds, during their main sequence evolution and afterward.  The observational and theoretical status of these winds is somewhat uncertain, due to the effects of line blanketing (which affects the mapping between spectral type and effective temperature; \citealt{martins02a, martins05a, repolust04a}) and of wind clumpiness (which affects the inferred momentum flux).   Nevertheless the wind-luminosity relation \citep{kudritzki99a}, which appears to remain valid after these corrections \citep{repolust04a}, implies that a young cluster massive enough to sample the entire initial mass function emits about one half as much force in winds as in starlight, $\dot{M}_w v_w = f_w L/c$ with $f_w\approx 0.5$.   The ratio $f_w$ reflects the flux-averaged optical depth of line overlap in the winds from hot stars, and so indicates a transfer of momentum from photons to wind.\footnote{In optically thick Wolf-Rayet winds and LBV outbursts, by contrast, the wind momentum can exceed $L/c$; however the stellar energy budget is smaller after core H burning, and we assume the momentum budget is as well.} 

In a dense cluster, the winds emitted by different stars rapidly collide with each other and shock-heat up to temperatures of $\sim 10^7$ K. The hot, shocked gas will then accelerate to supersonic speeds in a thermally-driven wind \citep{canto00a}, creating a bubble in the surrounding $\sim 10^4$ K gas. Models of stellar wind bubbles \citep{castor75a, koo92a, koo92b} typically assume that motions are efficiently energy-driven, rather than momentum-driven, so long as the wind is fast enough for its stopping shock to be non-radiative. An energy-driven bubble expands much more violently than the momentum-driven solutions presented above: the kinetic energy of a radiation-driven shell is $\riidot/(f_w v_w)\sim 10^{-2}$ times smaller than the wind energy. Thus our solution is valid only if the expansion is momentum-driven. However, \citet{harper-clark09a} point out that wind energy is easily lost if the shell is porous so the shocked wind gas can escape. Indeed, such escape is mandated by observations showing that the x-ray luminosity of large bubbles is far below what would be expected if all of the stellar wind energy remained trapped.

We use a modified version of \citeauthor{harper-clark09a}'s analysis to quantitatively estimate what this implies for $f_{\rm trap,w}$. Suppose that the shell of material swept up the by the expanding \hii region covers a fraction $C_f$ of the sky as seen from the driving cluster; for blister case clearly $C_f \leq 1/2$, while for the embedded case $C_f$ could take any value between 0 and 1 depending on the shell's porosity. Stars in the cluster launch winds with a mass flux $\dot{M}_w$ at a velocity $v_w$, and the fast wind material shock-heats to a temperature $T_X \gg T_{\rm II}$. The hot gas exerts a pressure $P_X = \rho_X c_X^2$ on the shell, where $\rho_X$ is the density of the shocked material within the shell and $c_X$ is its 
isothermal
sound speed, so
\begin{equation}
\label{ftrapw}
f_{\rm trap,w} = \frac{\rho_X c_X^2}{L/(4\pi \rii^2 c)}.
\end{equation}

The pressure of hot gas is set by the balance between the injection and escape of mass and thermal energy.  Both are injected directly by stellar winds and escape through holes in the shell, but additional mass is incorporated through the interaction of hot gas with the dense shell.   We assume this stripping occurs through hydrodynamical ablation rather than thermal evaporation, partly because magnetic fields inhibit thermal conduction.  Similarly, stellar photons only heat gas to temperature $\tii$, not to $T_X$, so they do not contribute; heating of gas to $T_X$ by x-ray photons is relatively insignificant.  The equations for mass and energy conservation of the hot gas are 
\begin{eqnarray}
\label{HotGasConservation1}
\dot{M}_X &  = & \dot{M}_w - 4\pi \rii^2 (1-C_f) \rho_X c_X + \dot{M}_{\rm abl} \\
\dot{E}_X & = & L_w - 4\pi \rii^2 (1-C_f) \left(\frac{5}{2}\rho_X c_X^3\right) \nonumber \\
& & \quad {}- 4\pi\rii^2 C_f P_X \riidot.
\label{HotGasConservation2}
\end{eqnarray}
where middle terms in both equations assume the hot fluid escapes through holes at speed $c_X$ and has its average properties as it does so.  Leakage ensures that mass and energy do not accumulate, implying that the LHS of both equations is effectively zero; furthermore the last term in $\dot{E}_w$, which represents adiabatic work, is negligible in the presence of holes because $C_f \riidot \ll (1-C_f) c_X$. 


The term $\dot{M}_{\rm abl}$ is the rate at which ionized gas at temperature $\tii$ ablates off the inner edge of the shell and heats up to $T_X$ through its interaction with the wind. The ablation process is a complex one, and the rate of ablation depends on the thermal behavior of the shocked wind. When the cooling time 
within the mixing layer 
is long, as it 
tends to be when the hot gas originates in a fast stellar wind.
mass ablation occurs in the ``jet-limited" regime described by \citet{canto91a}. In this case the ablation rate is limited by the rate at which the wind can supply momentum to a thin mixing layer between the two fluids, and 
these authors estimate that 
the ablation rate per unit area is $0.09 \rho_{\rm II} \cii^2 / (2 c_X)$, where $\rho_{\rm II}$ is the density of the 
ionized layer abutting the mixing layer. 
Assuming this process occurs over the entire shell, the total ablation rate is
\begin{equation}
\label{mablation}
\dot{M}_{\rm abl} = 4\pi \rii^2 
C_f
 \left(0.09 \rho_{\rm II} \frac{\cii^2}{2 c_X}\right).
\end{equation}
Since the ionized gas is in pressure balance with the hot gas, 
$\rho_{\rm II} \cii^2 =\rho_X c_X^2$.   Inserting this and $L_w = \dot{M}_w v_w^2/2$ into equations (\ref{HotGasConservation1}) and (\ref{HotGasConservation2}) and omitting the negligible terms, 
\begin{eqnarray}
\dot{M}_w & = & 4\pi\rii^2 \rho_X c_X \left[(1-C_f)-0.045 C_f\right] \\
\dot{M}_w v_w^2 & = & 20 \pi \rii^2 \rho_X c_X^3 (1-C_f).
\end{eqnarray}
The solution is  
\begin{equation} \label{P_X} 
\rho_X c_X^2 = {\dot{M}_w v_w/ ( 4\pi \rii^2)  \over  \left[ 5 (1-C_f)(1-1.045 C_f)\right]^{1/2} } 
\end{equation} 
so that, using $(1-C_f)(1-1.045 C_f)\simeq (1-1.02 C_f)^2$, 
\begin{equation}
f_{\rm trap,w} \simeq {f_w\over \sqrt{5} (1-1.02 C_f)}.
\end{equation} 
The divergence as $C_f$ approaches unity is not real, as our neglect of adiabatic losses and of accumulation within the shell are incorrect when the holes close up.   Likewise values of $f_{\rm trap,w}$ less than $f_w$ are not realistic, because the wind force is always present; our error in this limit is to assume that the wind energy is thermalized, when in fact it remains mostly kinetic if the shell is mostly holes.  


Since $f_w\simeq 0.5$,
 this implies that $f_{\rm trap,w}\simeq 0.22/(1-C_f) \sim 1$ unless we are considering the case of an embedded \hii region with an extremely non-porous shell, $1-C_f \ll 1$. Such low porosity is implausible given the turbulent, clumpy nature of the interstellar medium
 and the fact that pressure-driven shocks tend to run down density gradients and ``blow out''.
 Even if one started with a perfectly uniform ISM, the expanding shell is subject to \citet{Vishniac83a} instability of a pressure-driven slab; moreover if the wind caused the expanding shell to accelerate then 
the Rayleigh-Taylor instability would
spontaneously create holes in the shell, reducing $C_f$ below unity. Thus we conclude that $f_{\rm trap,w}$ is at most a few, and is likely to be small than order unity. Our conclusion is consistent with the numerical simulations of \citet{tenorio-tagle06a}, who also find that, in a non-uniform medium, the bulk of the mass around a young star cluster is not swept up into the thermal wind driven by $\sim 10^7$ K gas. Instead, that gas escapes rapidly through the porous shell, while the bulk of the mass expands more slowly (see their Figure 9).

The above calculation depends somewhat on our estimate of the term $\dot{M}_{\rm abl}$, which is uncertain for several reasons: because the physics of ablation is not well understood,  because the ablating area could be very different from $4\pi C_f \rii^2$, for instance if the shell's structure is more interesting than a broken sphere, and because other mechanisms like thermal evaporation, photoevaporation, and cloud disruption can all inject mass. Similarly, the density of the ablating gas could be higher than indicated by our pressure-balance argument, since photoevaporation might compress the gas \citep[e.g.][]{bertoldi90}. Given these uncertainties, one might consider $c_X$ better constrained than $\dot{M}_{\rm abl}$ -- for instance, an upper limit on the X-ray luminosity implies a lower limit on $c_X$.   Taking $c_X$ as given and solving equations (\ref{HotGasConservation1}) and (\ref{HotGasConservation2}) for $P_X$, we find 
\begin{equation}\label{ftrapw_from_cX}
f_{\rm trap,w} \simeq {f_w\over 5(1-C_f)} {v_w\over c_X}, 
\end{equation}
so trapping is important when $(1-C_f) c_X/v_w < f_w/5 \simeq 0.1$. 

Protostellar winds represent a separate, brief, but very intense phase which deserve separate mention.  Being magnetically launched, they are strongly collimated and may far exceed the photon momentum; moreover the entire stellar population generates them.   Rather than assess them directly we appeal to the treatment by \citet{matzner07}, who found that protostellar winds are very significant in the formation of the Pleiades and Orion Nebula clusters, but quite insignificant in the formation of the Arches or more massive clusters.   

\subsection{Trapped Infrared Photons}\label{trappedphotons} 

If the expanding shell traps the infrared light emitted within it, then its momentum will exceed that of the driving starlight by a factor 
\begin{equation}
\label{ftrapir}
\ftrapIR =  \frac{P_{\rm IR}}{L/(4\pi \rii^2 c)},
\end{equation}
where $P_{\rm IR}$ is the pressure of the trapped IR radiation field inside the shell.  We first consider the highly idealized case of a uniform, non-porous shell, $C_f = 1$, which provides an upper limit on $\ftrapIR$, before treating leakage through holes.   

When a uniform shell is optically thick to its own thermal emission, it radiates from a photosphere above which the flux-averaged optical depth is 2/3.  The emission is characterized by the shell's effective temperature, which satisfies $4\pi \rii^2 \sigma_{\rm SB} \Teffsh^4 = L$, and the flux-averaged mass opacity above the photosphere is approximately the Planck mean $\kappa_P(\Teffsh)$, implying a column $\Sigma_{\rm ph}\simeq2/[3\kappa_P(\Teffsh)]$ above the photosphere.   Shells with $\Sigma_{\rm sh} = \msh/(4\pi \rii^2)<\Sigma_{\rm ph}$ are optically thin to reprocessed light.  Thick shells, with $\Sigma_{\rm sh}\gg \Sigma_{\rm ph}$, can be treated in the diffusion approximation ($d P_{\rm IR}/\kappa_R(T) = - \sigma_{\rm SB} \Teffsh^4 d\Sigma$ with $P_{\rm IR}=a T(\Sigma)^4/3$ and $\kappa_R$ the Rosseland mean mass opacity), below the photosphere.  The solution for the pressure within an optically thick shell is  
\begin{equation}\label{Prad} 
P_{\rm IR} \simeq {\cal F}^{(-1)} \left[ {\cal F}\left( {a \Teffsh^4\over 3}\right) + \Sigma_{\rm sh}-\Sigma_{\rm ph}\right], 
\end{equation} 
where ${\cal F}(P_{\rm IR}) = \int_0^{P_{\rm IR}} dP_{\rm IR}'/\kappa_R(T')$, and ${\cal F}^{(-1)}$ is the inverse function of $\cal F$.   Evaluating equation (\ref{Prad}) for the \citet{weingartner01} standard dust model ``A'' for $R_V=5.5$, we find that 
\begin{equation}\label{ftrapIR_ClosedShell} 
\ftrapIR \simeq \left[ \Sigma_{\rm sh}^{-3} \left(132\over\Teffsh\right)^{6} + \Sigma_{\rm sh}^{-1.92 }  \left(72\over\Teffsh\right)^{1.71} \right]^{-2/3},
\end{equation} 
to good accuracy, where $\Sigma_{\rm shell}$ is understood to be in g\,cm$^{-2}$.   This result shows that radiation trapping can be quite significant, for instance when $\Sigma_{\rm shell} \gtsim 1$ and $\Teffsh > 60$ K, as is expected around a luminous young cluster; however this estimate is unrealistically high when radiation can leak away. 

In the more realistic case where leakage is important, we can compute $P_{\rm IR}$ in a manner analogous to our calculation of $P_X$, by balancing the rate at which energy the stars inside the shell add energy against the rate at which it leaks out. Our treatment here is a simplified version of that given in Appendix D of \citet{mckee08a}. We limit our attention to the case where the shell is optically thick ($\Sigma_{\rm shell}\gg \Sigma_{\rm ph}(\Teffsh)$) on average: in this limit we can compute the energy density of the trapped radiation field when the shell is porous, $C_f < 1$, simply by treating the shell as a perfectly opaque sphere with holes in it.  In this case energy balance requires that
\begin{equation}
L = (1-C_f)L + 3\pi \rii^2 (1-C_f) P_{\rm IR}.
\end{equation}
Here the left hand side is the rate at which the stars in the cluster inject radiant energy. The first term on the right hand side is the rate at which this beamed radiation field escapes the shell without interacting, while the second term represents the rate at which the reprocessed infrared radiation field leaks out. In writing this term, we have approximated that the IR radiation is isotropic, so the radiation flux escaping through a hole in the shell is given by the expression for the flux through an infinitesimal pinhole in an oven: $F =  3c P_{\rm IR}/4$. If the holes have finite size the true flux will be higher because the radiation field in the vicinity of the hole will be beamed outward. Solving this equation for $P_{\rm IR}$ gives
\begin{equation}
P_{\rm IR} = \frac43 \left(\frac{C_f}{1-C_f}\right) \frac{L}{\pi \rii^2 c},
\end{equation}
and substituting this into equation (\ref{ftrapir}) gives
\begin{equation}
\ftrapIR =  \frac43 \left( C_f\over1-C_f\right) 
\end{equation}
As with $f_{\rm trap,w}$, this is an upper limit. The true value will be smaller because radiation leaks out of the shell by diffusion at a finite rate even when $C_f=1$, because we have assumed that the internal radiation field has had an infinite amount of time to reach its steady state value, and because we have neglected the enhanced rate of energy loss through holes of finite size compared to those of infinitesimal size.

Even with this overestimate, our analysis shows find that for realistic values of $C_f \la 1/2$, the ratio of the force provided by trapped IR to that provided by the direct radiation field is $\ftrapIR \la 1$. We further note that, even if we started with a perfectly uniform ISM and did not have any winds to punch holes in the shell, numerical simulations show that a trapped IR radiation field is in itself sufficient to induce a Rayleigh-Taylor-like instability that punches holes in the shell \citep{krumholz09c}. This will ensure that $C_f$ is always well below 1, so that $\ftrapIR$ is no more than a few.

\subsection{Trapped Lyman $\alpha$ Photons}

Even blown-out sections of the shell that are transparent to infrared radiation may still be opaque to Lyman $\alpha$ photons, since they interact via a resonant process that produces a far larger cross section than absorption of IR photons by dust grains. As a result, our analysis based on $C_f$ does not apply to Ly $\alpha$ radiation. However, as pointed out by \citet{henney98a}, the influence of Ly $\alpha$ trapping is severely limited by the presence of dust grains. Absorption by dust grains in the \hii region interior or its bounding shell converts Ly $\alpha$ radiation to IR, which can then escape through the porous shell as discussed in the previous section. This process limits the energy density and pressure that can build up in Lyman $\alpha$. We refer readers to \citeauthor{henney98a} for a detailed treatment of this problem, and here simply quote the result: the Lyman $\alpha$ radiation pressure saturates at a value $P_{\rm Ly\alpha} \approx 0.06 \rhoii \cii^2 \sigma_{\rm d,0}^{-1}$, where $\sigma_{\rm d,0} = \sigma_{\rm d}/5\times 10^{-22}\mbox{ cm}^{-2}$ and $\sigma_{\rm d}$ is the dust cross section per H nucleus for Ly $\alpha$ photons. Milky Way gas has $\sigma_{\rm d,0}\approx 1$. We therefore have
\begin{equation}
f_{\rm trap,Ly\alpha} = \frac{P_{\rm Ly,\alpha}}{L/(4\pi\rii^2 c)} = 0.06 \sigma_{\rm d,0}^{-1} \xii^{1/2}.
\end{equation}
Thus during the radiation-pressure dominated phase, $\xii < 1$, trapped Ly $\alpha$ radiation is negligible in comparison to direct radiation pressure unless the dust is less opaque than typical Milky Way gas by a factor of $\ga 20$. Once gas pressure begins to dominate, $\xii > 1$, trapped Ly $\alpha$ radiation may exert more pressure than direct stellar radiation, but in this case it will still be negligible in comparison to gas pressure. For this reason we can simply set $f_{\rm trap,Ly\alpha} \approx 0$ without making a significant error in any phase of the expansion.

\section{Radiation-Driven \hii Regions Versus Supernovae}
\label{supernovae}

We have argued that radiation pressure dominates the dynamics of \hii regions in massive clusters, but this by itself does not establish that radiation-dominated \hii regions are the dominant source of feedback in star clusters. The primary alternative model is that supernovae are the main form of feedback \citep[e.g.][]{parmentier08a, baumgardt08a}. However, we can show on the basis of time scales that supernovae can play only a limited role. The crossing time in an object (either star cluster or gas cloud) of mass $M$ and surface density $\Sigma$ is \citep{tan06a}
\begin{equation}
t_{\rm cr} = \frac{0.95}{\sqrt{\alpha_{\rm vir} G}}\left(\frac{M}{\Sigma^3}\right)^{1/4} = 0.25 \alpha_{\rm vir}^{-1/2} M_4^{1/4} \Sigma_0^{-3/4}\mbox{ Myr},
\end{equation}
where $\alpha_{\rm vir}$ is the object's virial ratio \citep{bertoldi92}, $M_4=M/(10^4\,\msun)$, and $\Sigma_0=\Sigma/(1\mbox{ g cm}^{-2})$. Both observations and theory indicate that gas clumps that are forming massive star clusters have $\Sigma_0\gtsim 1$ \citep{mckee03, krumholz08a}, and a virial ratio $\gg 1$ would preclude the possibility of forming a gravitationally bound cluster. Thus we expect $t_{\rm cr}\sim 10^5$ yr unless the cluster mass exceeds $10^6$ $\msun$. 
This theoretical argument is in good agreement with observed star cluster crossing times in cases where those can be measured directly. For example, \citet{mccrady07a} find crossing times of $\sim 10^5$ yr for 21 clusters in M82. Thus we conclude that the crossing time in protoclusters is at least an order of magnitude shorter than the $\approx 3$ Myr required for the first supernova explosion. Numerical studies of star cluster formation show that, in the absence of a source of energy to drive the turbulence (such as stellar feedback), star formation generally proceeds to efficiencies of order unity within a few crossing times \citep[e.g.][]{bonnell03}. Thus supernovae could be an effective form of feedback only if some other, different energy source, most likely a different form of stellar feedback, inhibited star formation and kept the protocluster from collapsing for $\sim 10$ crossing times. While there is some observational evidence for such an extended star formation history \citep{tan06a, huff06}, even then supernovae would at most remove the last vestiges of gas from the cluster after most of the gas had been consumed by star formation or expelled by other mechanisms.

This argument argument does not by itself rule out the possibility that supernovae from neighboring star clusters that formed earlier might act as a source of delayed feedback. However, this idea also has timescale problems. In Orion, for example, there have been only $10-20$ SN explosions in the last 12 Myr \citep{bally08a}, a rate of $1-2$ per Myr. Thus the typical cluster-forming cloud is not hit by a SN blast wave within its $0.1$ Myr crossing time. Moreover, even if a blast wave did hit a cluster-forming cloud, any shock strong enough to affect a cloud significantly tends to destroy it completely, rather than leaving it intact but exciting turbulent motions \citep{nakamura06a}. Thus if SN from neighboring clusters were a dominant form of feedback, then most cluster-forming gas cloud would their lives being shredded by a blast wave. This is inconsistent with the observed morphologies of disrupting clouds, which typically show no evidence for interaction with a strong external shock.

\section{Summary}
\label{summary}

We examine the role of radiation pressure in determining the dynamics of expanding \hii regions around young star clusters. We derive a characteristic number (equation \ref{zeta}) that describes the importance of radiation pressure in \hii region expansion, and we show that for very massive star clusters such as the Arches or the large clusters found in starburst galaxies, this number is always $\gg 1$, indicating that radiation pressure dominates at least the initial phases of expansion. For such clusters, we derive a solution describing the expansion of a shell of material driven by a combination of radiation and gas pressure. Our solution shows that radiation pressure is capable of driving expansion at speeds considerably larger than the ionized gas sound speed, and therefore radiation-driven \hii regions can be an effective feedback mechanism in clusters where a high escape speed renders gas-driven \hii regions ineffectual.

Radiation-driven \hii regions are likely to be the dominant feedback mechanism that regulates the formation of massive star clusters. Unlike supernovae, they begin to operate immediately upon formation of the stars, rather than with a delay of many dynamical times. While stellar wind bubbles or trapped infrared radiation fields are potentially more effective than direct stellar photons as a source of feedback, we argue that they are unlikely to be dominant because they are sapped by leakage of energy out of the expanding shell. Future models of feedback in massive star cluster formation must therefore be careful to take radiation pressure-driven \hii regions into account.

\acknowledgements We thank S.~M.\ Fall, J.~R.\ Graham, C.~F.\ McKee, and E.~Ramirez-Ruiz for helpful discussions. We thank the referee, W.\ Henney, for a report that increased the quality of the paper. Support for this work was provided by the Alfred P.\ Sloan Foundation (MRK), an Ontario Early Researcher Award (CDM), by NASA, as part of the Spitzer Theoretical Research Program, through a contract issued by the JPL (MRK and CDM), by NSERC Canada (CDM), and by the National Science Foundation through grant AST-0807739 (MRK).  

\appendix

\section{The Role of Dust Absorption}
\label{dustabsorption}

For most of the time that an \hii region is expanding, dust within it will be able to absorb photons. Assuming that all of a source's radiant energy is eventually absorbed by dust grains (even if ionizing photons are first absorbed by hydrogen atoms and are eventually down-converted to photons below the Lyman limit), the equilibrium dust temperature at a distance $r$ from the source is given implicitly by
\begin{equation}
\frac{L}{4\pi r^2} = \int Q^A B_{\nu}(T)\, d\nu,
\end{equation}
where $Q^A$ is dust the absorption efficiency, which depends on grain properties. Dust can survive wherever $T$ is below the sublimation temperature $T_{\rm s}$; for the most refractory grains \citet{wolfire86a, wolfire87} estimate $\int Q^A B_{\nu}(T_{\rm s})\, d\nu = 1.24\times 10^7$ erg cm$^{-2}$ s$^{-1}$, so at least some dust exists at all radii greater than the sublimation radius
\begin{equation}
r_{\rm s} = 3.8\times 10^{-4} \psi^{1/2} S_{49}^{1/2}\mbox{ pc}. 
\end{equation}
Even for the most luminous sources we consider, $S_{49}\sim 10^4$, this is much less than the cluster radius, $r_{\rm ch}$, or $r_{\rm stall}$. However, dust absorption does not affect the dynamics of the radiation pressure-dominated phase directly, since the dust and gas are likely to be well-coupled by magnetic fields. The momentum transferred by radiation to the gas therefore does not depend on whether photons are absorbed by dust grains or hydrogen atoms. However, dust grains do play a role when gas pressure becomes important, because they modify the ionized gas density $\rhoii$ and thus change the gas pressure; this dependence is captured through our parameter $\phi$. A full discussion of the behavior of dust inside \hii regions is beyond the scope of this work, and we refer readers to \citet{dopita02a} and \citet{arthur04a}. In this appendix we make some simple estimates to show that our use of a constant $\phi$ is unlikely to lead to serious error.

\citet{arthur04a} show that the value of $\phi$ (which is equivalent to the parameter $y^3$ in \citeauthor{arthur04a}'s treatment) is determined by the dust optical depth $\tau_d$ through the ionized gas; for typical gas pressure-dominated \hii regions whose metallicity is not vastly greater than Solar, $\tau_d\la 1$ and $\phi \ga 0.5$ is appropriate, consistent with our value of $0.73$ from \citet{mckee97}. To see how this might vary for the radiation-dominated case, it is useful to estimate $\tau_d$ in terms of the ionization parameter
\begin{equation}
\mathcal{U} \equiv \frac{F}{cn},
\end{equation}
where $F$ is the ionizing photon flux into the recombining gas and $n$ is its density. Ionization-recombination balance requires $\phi F = \alphab n^2 h$, where $h$ is the thickness of the recombining gas layer and $n$ is its number density, so the dust optical depth is
\begin{equation}
\tau_d = n h \sigma_d = \frac{\phi c \sigma_d}{\alphab} \mathcal{U} \simeq 90 \phi \mathcal{U},
\end{equation}
where $\sigma_d$ is the dust cross section per H nucleus to ionizing photons. The numerical evaluation uses $\sigma_d \simeq 1.0\times 10^{-21}$ cm$^{2}$, appropriate for Solar metallicity.

Since $\phi \leq 1$, this result shows that $\tau_d \la 1$ as long as $\mathcal{U} \la 0.01$. When gas pressure dominates, $n$ is constant and an increase in ionizing luminosity increases $\mathcal{U}$. However, once radiation pressure becomes significant, there is a countervailing effect: radiation pressure piles up gas against the inner wall of the expanding shell, increasing its density and thus driving $\mathcal{U}$ down. Under the assumption of hydrostatic equilibrium, \citet{dopita02a} find that this effect ensures that $\mathcal{U}$ never rises above $\sim 0.01$. While our expanding shell is of course not in hydrostatic equilibrium, as long as the expansion is slow compared to the signal speed within the shell, quasi-hydrostatic equilibrium should be established in the shell interior, and thus we also expect a maximum value of $\mathcal{U}\sim 0.01$. This ensures that dust absorption never dominates over ionized gas absorption, and we can safely adopt a constant, order-unity value of $\phi$.

However, we note that even if this argument is incorrect, we only make a significant error in following the dynamics when $\xii\sim 1$, i.e.\ when gas and radiation pressure are nearly of equal importance. That is because our choice $\phi=0.73$ is appropriate when gas pressure dominates ($\xii \gg 1$), and does not affect the dynamics significantly when radiation pressure dominates ($\xii \ll 1$). Since our treatment of the crossover regime is very approximate in any event, inaccuracy in estimating $\phi$ in this regime would not further degrade our accuracy.

\bibliographystyle{apj}
\bibliography{refs}

\end{document}